\documentclass[12pt]{article}
\usepackage{graphicx}

\title{Elliptic flow orientation, saturation and low $p_t$ behavior}
\author{V.A. Abramovsky\thanks{email: ava@novsu.ac.ru},  A.V. Popov\thanks{email: avp@novgorod.net} \\
Novgorod State University, B. S.-Peterburgskaya Street 41,\\
Novgorod the Great, Russia, 173003}

\begin{document}
\maketitle
\abstract{We consider general model with factorization between macroscopic flow and matter decay distribution. 
We show universality of $p_t^2$ behavior of elliptic flow at small $p_t$ for identified final particles in the symmetric nucleus collision.
At high $p_t$ we compare non-relativistic and relativistic models for boosted decay distribution. In the relativistic models
with distribution having power-like tail we show existence of elliptic flow saturation. This means that the elliptic flow value $v_2$ tend to constant
at high $p_t$. We discuss the importance of determination of elliptic flow 
orientation which can help us to compare different models. For example, we introduce potential expansion model which can reproduce azimuthal asymmetry 
but based on the assumption that there are only cold strong forces and no thermalization. 
This model show opposite sign of elliptic flow orientation in comparison with thermal model. 
We also show that at small $p_t$ sign of elliptic flow can change due to sign of second derivative of decay distribution.
This fact tell us that we must more carefully link elliptic flow sign and properties of the model.}

\section{Introduction}
Azimuthal momentum space distributions is very interesting effect widely studying at RHIC. The strong elliptic flow
generated in non-central collisions was founded in the set of experiments. Basis method 
is study Fourier expansion of azimuthal particle distributions \cite{Voloshin_94}. For study elliptic flow we
interesting $v_2$ which is calculated from azimuthal average $\langle \cos(2\phi) \rangle$, where angle $\phi$
defined relatively to reaction plain which is must be determined for each event and defined as plain based on the beam axis and line joining center of the nucleous. Next, we must perform average over all
set of events.

There are many phenomenological explanation of the elliptic flow phenomenon. The main goal of the theoretical studying
is explain how transverse plain asymmetry of matter in the reaction zone is transformed into asymmetry of the momentum space
distribution. Existence of the azimuthal asymmetry is striking evidence of reinteractions in the reaction zone which enforce
matter into asymmetric radial expansion.
The modern most desired explanation of low $p_t$ elliptic flow is existence of the Quark Gluon Plasma (QGP) and thermalization stage during nucleus collision.
This allow us to use hydrodynamic with relativistic equation of motion $\partial_\mu T^{\mu\nu}$. But it is necessary to know pressure as function of 
energy density and chemical potential. Pressure can be founded from QGP equation of state which unknown, especially near phase transition point. 
Many numeric calculations performed in the longitudinal boost invariant ideal fluid model with equation of state of ideal gas: $P\sim T^4$. 
In additional dissipative effects, non-equilibrium transport equations and other improvements can be used with hydrodynamic model.
Unfortunaly, we can not reach full physical understanding because too much uncontrolled phenomenological assumptions but so small 
experimental observable variables. 

Recent experimental data from RHIC for Au+Au collisions for identified particles  \cite{Star:v_2_id} show saturation of $v_2$ in the region $p_t>2 GeV$, Fig. \ref{Fig_v2_main}.
Hydrodynamic calculation can not successful reproduce $v_2$ behavior in the saturation region. The origin saturation is still unknown.
At high $p_t$ was proposed a mechanism of jet quenching -- high $p_t$ pertrubative gluons propagated through asymmetric reaction zone and
lost its momentum interacting with dense matter \cite{Wang_00}. Also proposer the contribution to $v_2$ from nucleus classical fields in the saturation regime \cite{Saturation_02}. 
In our calculation we show that the saturation phenomenon have mostly kinematic origin, but saturation values depend on details of
matter motion due to reinteractions. 

In this paper we try to separate contributions to $v_2$ from macroscopic flow of matter $u(\vec \beta)$ and from effect of the local matter decay. And try
to describe low $p_t$ and high $p_t$ of $v_2$ behavior in unified language.
In our model for final particles spectrum we assume factorization between macroscopic flow and matter decay distribution $f$. In general case,
$u(\vec \beta)$ can be hydrodynamic flow and $f$ be a thermal distribution of specified hardron at the freeze-out temperature. So we can say
that we add local decay matter effects to the hydrodynamic model. 
On the other hand, $u(\vec \beta)$ can be color string transverse motion due its repulsion and $f$ be standard color string fusion spectrum.
In same sense we generalize Blast Wave model \cite{Voloshin_02} and we not concerning on real sense of flow $u(\vec \beta)$ and decay distribution $f$.
The model does not cover jet quenching because pertrubative jets produced at the early stage of collision where macroscopic flow is still not formed.



\section{Elliptic flow at low and high $p_t$}

Consider elliptic flow in c.m. system at at midrapidity, $y=0$. Assuming boost invariance as usual we may reduce problem to the 
two dimension transverse plain calculations.  
Let $u(\vec \beta)=u(\beta,\psi)$ be a density of expanding matter depend on transverse flow velocity $\vec \beta$. We always assume
that exists some $\beta_{max}<1$ and $u(\vec \beta)=0$ if $|\vec \beta|>\beta_{max}$.
For final transverse particle distribution we can write
    \begin{equation}\label{F_def}
     F(\vec p)=F(p,\phi)=\frac{1}{Z(p)}\frac{d^2 N}{dp^2}=\frac{1}{Z(p)}\int u(\vec \beta)f_\beta(\vec p) d^2\beta
    \end{equation}
where $f(\vec p)=f(|p|^2)$ -- transverse fusion spectrum of small piece of expanding matter,
$f_\beta(\vec p)$ -- modified fusion spectrum of fixed particle type according to it transverse motion with $\vec\beta$ velocity (one can speak about "blue shift"). 
With aim to exclude final spectrum mixing between particles of different species we consider only  elliptic flow for identified particles.
Factor $Z(p)$ defined to satisfy azimuthal normalization
    \begin{equation}
    \int\limits_0^{2\pi}F(p,\phi)d\phi=1
    \end{equation}
    \begin{equation}
    Z(p)=\int \left(\int u(\vec \beta)f_\beta(\vec p) d^2\beta\right) d\phi
    \end{equation}

\begin{figure} 
  \centering
  \includegraphics[width=1\textwidth]{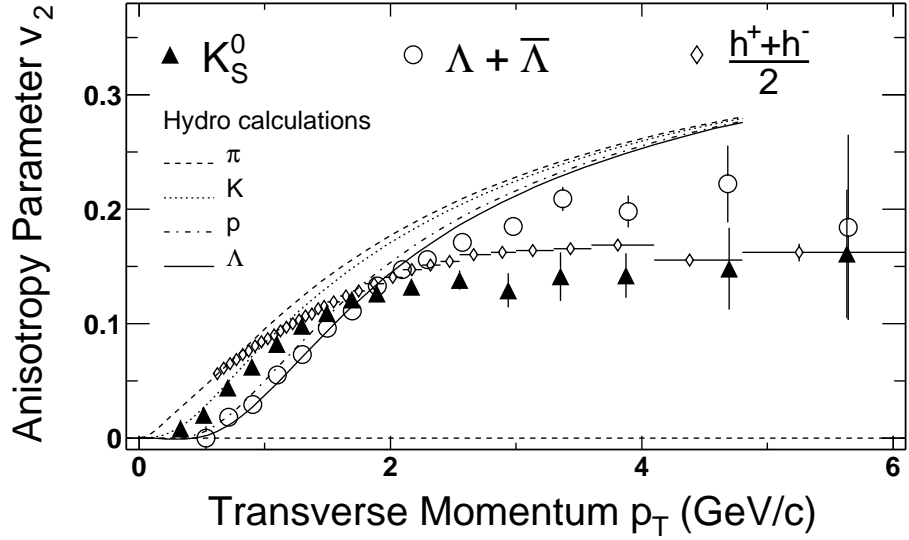}
  \caption{The minimum-bias (0--80\% of the collision cross section) $v_{2}(p_T)$ for $K_{S}^{0}$, $\Lambda +
\overline{\Lambda}$ and $h^{\pm}$~\cite{Star:v_2_id}.} \label{Fig_v2_main}
\end{figure}
 
We consider only collisions of equal nucleus such as \mbox{Au+Au}.
Then the distribution $u(\vec\beta)$ have very important symmetry properties which follow from the basic 
properties of the nucleus collision geometry.
    \begin{equation} \label{u_prop1}
    u(\beta,-\psi)=u(\beta,\psi+\pi)=u(\beta,\psi)  
    \end{equation}
This automatic gives
    \begin{eqnarray}
    \int_0^{2\pi} u(\beta,\psi) \sin(n\psi) d\psi=0 \label{u_prop_sin}\\
    \int_0^{2\pi} u(\beta,\psi) \cos((2n+1)\psi) d\psi=0 \label{u_prop_cos}
    \end{eqnarray}
So for $u(\vec\beta)$ we have only even $\cos$ harmonics in the azimuthal Fourier decomposition. 

Elliptic flow can be calculated from Fourier decomposition of $F(p,\phi)$ 
    \begin{equation}\label{v2_def}
    v_2=\int\limits_0^{2\pi} F(p,\phi) \cos(2\phi) d\phi=\langle \cos(2\phi) \rangle
    \end{equation}
    \begin{equation}
    F(p,\phi)=\frac{1}{2\pi} \left( 1+2v_2\cos(2\phi)+\ldots\right)
    \end{equation}  

Now our task is to calculate $v_2$ from (\ref{v2_def}) using (\ref{F_def}).

The integrated $v_4$ is about a factor of 10 smaller than $v_2$ \cite{Star:v_4}.
In cases when $v_1$(directed flow)á $v_4$ and other higher harmonics enough small  
relatively to $v_2$ (one can say that the azimuthal 
distribution have shape like $\cos 2\phi$) we can simple estimate $v_2$ with sufficient accuracy 
    \begin{equation}
    v_2=\frac{\pi}{2}\left(F(p,0)-F(p,\frac{\pi}{2}) \right)
    \end{equation}

For find $f_\beta(\vec p)$ in relativistic form we must boost lab $\vec p$ to the comoving frame. Decompose
$\vec p$ into $\vec \beta$ and $\vec \beta$ orthogonal directions
    \begin{equation}
    \vec p=p\cos(\phi-\psi)\frac{\vec \beta}{|\beta|}+p\sin(\phi-\psi)\frac{\vec \beta_{ort}}{|\beta|}
    \end{equation} 
    
Now we can easy perform boost in $\vec \beta$ direction. In relativistic form for $f_\beta(\vec p)$ we have
    \begin{equation} \label{f_rel}
    f_\beta(\vec p)= f\left(p^2 \sin^2(\phi-\psi)+\frac{\left(p\cos(\phi-\psi)-\beta\sqrt{M^2+p^2}\right)^2}{1-\beta^2}\right) 
    \end{equation}

We also need to relativistic transform of phase space $d^{\:2}p$ in (\ref{F_def})
    \begin{equation}
    \frac{d^{\:2}p}{\sqrt{M^2+p^2}}=inv
    \end{equation}

Performing calculation we obtain    
    \begin{equation}
    F(p,\phi)=\frac{1}{Z(p)}\int u(\vec \beta)f_\beta(\vec p) \left(1-\frac{\beta p \cos(\phi-\psi)}{\sqrt{M^2+p^2}} \right) \frac{\beta d\beta d\psi}{\sqrt{1-\beta^2}}
    \end{equation}
where $M$ -- identified particle mass and distribution $f$ must be considered as function on $p^2$.

Consider $v_2$ behavior at small $p$. In this paper for simplicity we show calculation with $f_\beta(\vec p)$ in non-relativistic form
    \begin{equation} \label{f_beta}
    f_\beta(\vec p)=f(|\vec p - M\vec \beta|^2)=f(p^2+M^2\beta^2-2M\beta p\cos(\phi-\psi))
    \end{equation}
Relativistic calculation based on (\ref{f_rel}) also can be performed. The resulting $p^2$ behavior will not changed. Only
constant factor definition will changed.
    
Due to existence factor $Z^{-1}(p)$ in (\ref{F_def}), any $\phi$-constant multiplier 
in $\int u(\vec \beta)f_\beta(\vec p) d^2\beta$ does not contribute to $F(p,\phi)$ and can be dropped. Distribution
$f_\beta(p)$ at small $p$ can be expanded in Taylor series. Zero expansion order does not contribute to $v_2$ because there is
no dependence on $\phi$. In zero order $Z(p)$ does not vanish
    \begin{equation}
    Z(p)=Z_0=\int u(\beta,\psi) f(M^2 \beta^2) \beta d\beta d\psi
    \end{equation}

 In the first order we have $\int u(\psi)\cos(\phi-\psi)d\psi=0$ due to (\ref{u_prop_sin}) and (\ref{u_prop_cos}).
Second order have contribution to $v_2$
    \begin{equation}
    F(p,\phi)=\frac{1}{Z(p)}\int u(\beta,\psi) \frac{1}{2} f''(M^2\beta^2) (2\beta Mp)^2 \cos^2(\phi-\psi) \beta d\beta d\psi
    \end{equation}
where we omit all terms which give zero contribution to $v_2$.

If we need only $v_2$ then
    \begin{equation} \label{cos_2_phi_psi}
    \cos^2(\phi-\psi)=\cos^2\phi \cos^2\psi + \sin^2\phi \sin^2\psi=\frac{1}{2} \cos 2\phi \cos 2\psi
    \end{equation}

Performing all calculations we find
    \begin{equation} \label{v2_result}
    v_2=C\left(\{u\},\{f\},M\right) p^2
    \end{equation}
where
    \begin{equation} \label{v2_result_C}
    C=\frac{2\pi M^2\int u(\beta,\psi) \cos 2\psi f''(M^2\beta^2) \beta^3 d\beta d\psi}{\int u(\beta,\psi) f(M^2 \beta^2) \beta d\beta d\psi}
    \end{equation}
    
We found that the $v_2$ grow as $p^2$ at small p and vanish at $p=0$. This fact is not depend on detail form of $f(\vec p)$ distribution.
In the Fig. \ref{Fig_v2_main} we can see $p^2$-grow of $v_2$ for identified particles at small $p$.

Consider sign of $v_2$. In most cases $\int u(\beta,\psi) \cos 2\psi d\psi>0$ because we set zero azimuthal angle to flow maximum.
For any known distribution $f''(p^2)>0$ for $p>p_0$, where $p_0$ -- same specific sufficient small momentum. So we can conclude that 
in our model $v_2>0$ as observed\footnote{In the low energy collision mean $\beta$ can be small. So $f''(M^2\beta^2)$ and $v_2$ may be negative.}.

Now consider case when $p$ is large. For reason of explicit calculation we assume that the $f(p)$ at large $p$ have power like tail and 
have asymptotic $|p|^{-2\alpha}$. This is consistent with any known fits of particle distributions. At first we consider non-relativistic case.
    \begin{equation} \label{F_big_p}
    F=\frac{1}{Z(p)}\int u(\vec \beta) p^{-2\alpha}\left(1+\left(\frac{M\beta}{p}\right)^2-2\left(\frac{M\beta}{p}\right) \cos(\phi-\psi)\right)^{-\alpha} d^2\beta
    \end{equation}

We can use $M\beta/p$ as small value and expand (\ref{F_big_p}) in Taylor series.
    \begin{equation} \label{F_big_p_2}
    F=\frac{1}{Z(p)}\int u(\vec \beta) p^{-2\alpha}\left(1+2\alpha^2 \left(\frac{M\beta}{p}\right)^2\cos^2(\phi-\psi) \right) d^2\beta
    \end{equation}
where we omit first expansion order because it does not contribute to $v_2$.

In leading order we can calculate $Z(p)$ 
    \begin{equation}
    Z(p)= 2\pi p^{-2\alpha} \int u(\vec \beta) d^2\beta
    \end{equation}
    
For $\cos^2(\phi-\psi)$ in (\ref{F_big_p_2}) we can apply rule (\ref{cos_2_phi_psi}).Then we have
    \begin{equation}
    v_2=\left(\frac{M\beta\alpha}{p}\right)^2\frac{\int u(\vec \beta) \cos 2\psi d^2\beta}{2 \int u(\vec \beta) d^2\beta}
    \end{equation}
    
So we see $1/p^2$ decreasing of $v_2$ in the non-relativistic model.    

In the relativistic model we must determine asymptotic properties of (\ref{f_rel}) where $f\sim|p|^{-2\alpha}$.
    \begin{equation}
    f_\beta(\vec p)=p^{-2\alpha}\left|\frac{\beta^2}{1-\beta^2}\cos^2(\phi-\psi)-\frac{2\beta}{1-\beta^2}\cos(\phi-\psi)\right|^{-\alpha}
    \end{equation}

Inserting it into (\ref{F_def}) we see that the $|p|^{-2\alpha}$ factor absorbed into factor $Z(p)$ and does not influence on the $v_2$.    
So $v_2$ tend to some constant at high $p_t$ 
    \begin{equation}
    v_2= \frac{\int u(\vec \beta) R\left(\cos(\phi-\psi),\beta\right) cos(2\phi) d^2 \beta d\phi}{\int u(\vec \beta) R\left(\cos(\phi-\psi),\beta\right) d^2 \beta d\phi}
    \end{equation}
where $R\left(\cos(\phi-\psi),\beta\right)=f_\beta(\vec p)/p^{-2\alpha}$. 

It is interesting to know real $v_2$ behavior at high $p_t$ to distinguish relativistic and non-relativistic model. Unfortunaly,
the current experimental data have large error bars at $p_t>4 Gev$ and high $p_t$ behavior in not clear.

\section{Elliptic flow orientation and directed flow}
Asymmetric nucleus collision is collision between to different nucleus at midrapidity or collision of equivalent nucleus
at rapidity near $\pm y_{beam}$. If collision is asymmetric then (\ref{u_prop_cos}) for $n=0$ does not holds. And we can expect
existence of the directed flow $v_1\neq0$. Non-zero $v_1$ is usually linked with barion stopping, limited fragmentation and projectile nucleon bounce. 

Important property of directed flow is orientation relatively to reaction plain. Directed flow can lie only in the reaction plain due to
symmetry. Of course, we assume that there is no symmetry breaking in each nucleus collision. If $v_1$ directions was determined, the
elliptic flow $v_2$ must be calculated relatively this direction. If founded $v_2$ positive than we can conclude that the elliptic flow
is in reaction plain and maximum matter flow in direction where almond shape reaction zone is thin. And vice versa, if $v_2$ negative than 
elliptic flow in direction where reaction zone is thick.

Consider potential expansion model. In this model we assume that there are only potential forces acting on produced matter after
collision. Each two near pieces of matter after its production have positive potential energy of reinteraction. Instead abstract
"matter" we can think about strong color fields or about color strings. The origin of repulsive forces {\it ab initio} is not known. 
In pure phenomenological language we can address problem to non-pertrubative QCD vacuum properties. 

For simplicity we consider color strings placed in one dimensional line with equal spacing. Let point $x=0$ be a center of this 
string chain, $\rho$ -- string tension, $u$ - repulsive potential energy of neighboring string per unit length. After time evolution of this
system all potential energy transferred into kinetic energy of transverse string motion. Due to symmetry we can consider only
strings at $x>0$. Let index $i$ enumerate strings from $x=0$ to positive direction. String having $i=0$ after expansion will not moving due
to symmetry. Transverse rapidity $\theta_1$ of $i=1$ string 
    \begin{equation}
    \theta_1={\mathop{\rm arcch}\nolimits}\frac{u}{\rho}-1
    \end{equation}
Other strings rapidity is simple additive
    \begin{equation}
    \theta_i=i\theta_1
    \end{equation}
So longest string chain have larger transverse velocity of leftmost and rightmost strings and in thick direction of reaction zone we have
flow maximum.

Determination of $v_2$ sign relatively to $v_1$ orientation is very important for testing of validity of different models describing 
reinteractions during collision. For example, in the hydrodynamic-inspired models $v_2$ positive due to maximum pressure gradient in thin
direction. In the potential expansion model where role of matter play color strings we expect negative $v_2$ due to maximum energy released 
in thick direction. At different energies different mechanism can play a main role. First signal of $v_2$ positivity at RHIC energies
was found in \cite{Star:v_4}. But this is not full story. In our model we can see in (\ref{v2_result_C}) that $v_2$ sign can
changed not only due preferred expansion direction but due to negative sign of $f''$.    

Near midrapidity at small $y$ we can expand $u(\vec \beta,y)$ in Taylor series
    \begin{equation}
    u(\vec \beta,y)=u_0(\vec \beta)+u_1(\vec \beta)y+\ldots
    \end{equation}
where $u_0(\vec \beta)$ have zero directed flow and $u_1(\vec \beta)$ -- non-zero.

In our model directed flow have form
    \begin{equation}
    v_1(p)=\int\limits_0^{2\pi} \frac{1} {Z(p)} \left(\int u(\vec \beta) f_\beta(p) d^2\beta \right)\cos\phi d\phi
    \end{equation}

Consider $v_1$ at low $y$ and low $p$. Similarly (\ref{v2_result}) we get
    \begin{equation}
    v_1=Cyp
    \end{equation}  
where $C$ -- some constant not depended on $y$ and $p$. Note that the sign of $v_1$ have no sense due to impossibility of
its measurement. Such $y$ scaling we was observed in \cite{Star:v_4}. Dependence $v_1(p)$ is not measured yet. 

\section{Conclusion}
In this paper we have studied azimuthal anisotropy at large and small $p_t$ in generalized model with factorization between macroscopic flow and matter decay distribution.
We showed universality of $p^2$ behavior of elliptic flow at small $p$ for specified final particles --
$v_2$ grow as $p^2$ at small p and vanish at $p=0$. This fact is not depend on detail form of $f(\vec p)$ distribution and matter flow $u(\vec \beta)$.
So we can say that it is model independent.

At high $p_t$ we compared non-relativistic and relativistic models of Lorentz boosted distribution $f$. In the relativistic models
with distribution $f$ having power-like tail we show existence of elliptic flow saturation. 
In the non-relativistic model $v_2$ decrease as $1/p^2$. More precise measurements of $v_2$ at region $p_t>4 Gev$ are needed to 
make selection between models. 

Determination the sign of $v_2$ relatively to directed flow is important criteria to for testing of validity of different models 
describing reinteractions during collision. To make conclusion about model selection we need experimental determination of $v_2$ sign
in wide range of collision energy. For example, we can compare hydro and potential expansion models.

We also shown that at small $p_t$ sign of elliptic flow can change due to sign of second derivative of decay distribution $f$. This fact 
must be considered if we want to make model selection based on elliptic flow orientation. For example, for color string at sufficient
small $p_t$ we have $f''>0$.

\section*{Acknowledgments}
We thank A.V. Dmitriev and N.V. Prikhod'ko for useful discussions.
This work was supported by RFBR Grant RFBR-03-02-16157a and grant of Ministry for Education E02-3.1-282

\end{document}